\input harvmac\skip0=\baselineskip
\input epsf

\newcount\figno
\figno=0
\def\fig#1#2#3{
\par\begingroup\parindent=0pt\leftskip=1cm\rightskip=1cm\parindent=0pt
\baselineskip=11pt \global\advance\figno by 1 \midinsert
\epsfxsize=#3 \centerline{\epsfbox{#2}} \vskip 12pt {\bf Fig.\
\the\figno: } #1\par
\endinsert\endgroup\par
}
\def\figlabel#1{\xdef#1{\the\figno}}
\def\encadremath#1{\vbox{\hrule\hbox{\vrule\kern8pt\vbox{\kern8pt
\hbox{$\displaystyle #1$}\kern8pt} \kern8pt\vrule}\hrule}}



\lref\osv{H.~Ooguri, A.~Strominger and C.~Vafa,
  ``Black hole attractors and the topological string,''
  Phys.\ Rev.\ D {\bf 70}, 106007 (2004)
  [arXiv:hep-th/0405146].}

\lref\msw{
  J.~M.~Maldacena, A.~Strominger and E.~Witten,
  ``Black hole entropy in M-theory,''
  JHEP {\bf 9712}, 002 (1997)
  [arXiv:hep-th/9711053].
}

\lref\gv{
  R.~Gopakumar and C.~Vafa,
  ``M-theory and topological strings. II,''
  arXiv:hep-th/9812127.
}

\lref\DijkgraafXW{
  R.~Dijkgraaf, G.~W.~Moore, E.~Verlinde and H.~Verlinde,
  ``Elliptic genera of symmetric products and second quantized strings,''
  Commun.\ Math.\ Phys.\  {\bf 185}, 197 (1997)
  [arXiv:hep-th/9608096].
}

\lref\moorecharges{
  R.~Minasian and G.~W.~Moore,
  ``K-theory and Ramond-Ramond charge,''
  JHEP {\bf 9711}, 002 (1997)
  [arXiv:hep-th/9710230].
}

\lref\adstop{
  D.~Gaiotto, A.~Strominger and X.~Yin,
  ``From AdS(3)/CFT(2) to black holes / topological strings,''
  arXiv:hep-th/0602046.
}

\lref\freedwitten{
  D.~S.~Freed and E.~Witten,
  ``Anomalies in string theory with D-branes,''
  arXiv:hep-th/9907189.
}

\lref\moorebelov{
  D.~Belov and G.~W.~Moore,
  ``Holographic action for the self-dual field,''
  arXiv:hep-th/0605038.
}

\lref\fareytail{ R.~Dijkgraaf, J.~M.~Maldacena, G.~W.~Moore and
E.~P.~Verlinde,
  ``A black hole farey tail,''
  arXiv:hep-th/0005003.
}

\lref\marcus{
  A.~Klemm and M.~Marino,
  ``Counting BPS states on the Enriques Calabi-Yau,''
  arXiv:hep-th/0512227.
}

\lref\toroidal{
  J.~M.~Maldacena, G.~W.~Moore and A.~Strominger,
  ``Counting BPS black holes in toroidal type II string theory,''
  arXiv:hep-th/9903163.
}

\lref\meg{
  D.~Gaiotto, A.~Strominger and X.~Yin,
  ``The M5-brane elliptic genus: Modularity and BPS states,''
  arXiv:hep-th/0607010.
}

\lref\crazy{
  M.~x.~Huang, A.~Klemm and S.~Quackenbush,
  ``Topological string theory on compact Calabi-Yau: Modularity and boundary
  conditions,''
  arXiv:hep-th/0612125.
}

\lref\mms{
  J.~M.~Maldacena, G.~W.~Moore and A.~Strominger,
  ``Counting BPS black holes in toroidal type II string theory,''
  arXiv:hep-th/9903163.
}

\lref\Estring{
  J.~A.~Minahan, D.~Nemeschansky, C.~Vafa and N.~P.~Warner,
  ``E-strings and N = 4 topological Yang-Mills theories,''
  Nucl.\ Phys.\ B {\bf 527}, 581 (1998)
  [arXiv:hep-th/9802168].
}

\lref\deBoerVG{
  J.~de Boer, M.~C.~N.~Cheng, R.~Dijkgraaf, J.~Manschot and E.~Verlinde,
  ``A farey tail for attractor black holes,''
  JHEP {\bf 0611}, 024 (2006)
  [arXiv:hep-th/0608059].
}

\lref\KrausNB{
  P.~Kraus and F.~Larsen,
  ``Partition functions and elliptic genera from supergravity,''
  arXiv:hep-th/0607138.
}

\lref\DabholkarDT{
  A.~Dabholkar, F.~Denef, G.~W.~Moore and B.~Pioline,
  ``Precision counting of small black holes,''
  JHEP {\bf 0510}, 096 (2005)
  [arXiv:hep-th/0507014].
}

\lref\denefwork{
  F.~Denef and G.~W.~Moore, to appear.
}

\lref\kkv{
  S.~H.~Katz, A.~Klemm and C.~Vafa,
  ``M-theory, topological strings and spinning black holes,''
  Adv.\ Theor.\ Math.\ Phys.\  {\bf 3}, 1445 (1999)
  [arXiv:hep-th/9910181].
}

\lref\MinahanCT{
  J.~A.~Minahan, D.~Nemeschansky and N.~P.~Warner,
  ``Partition functions for BPS states of the non-critical E(8) string,''
  Adv.\ Theor.\ Math.\ Phys.\  {\bf 1}, 167 (1998)
  [arXiv:hep-th/9707149].
}

\lref\gvi{
  R.~Gopakumar and C.~Vafa,
  ``M-theory and topological strings. I,''
  arXiv:hep-th/9809187.
}

\lref\KlemmTX{
  A.~Klemm and S.~Theisen,
  ``Considerations of one modulus Calabi-Yau compactifications: Picard-Fuchs
  equations, Kahler potentials and mirror maps,''
  Nucl.\ Phys.\ B {\bf 389}, 153 (1993)
  [arXiv:hep-th/9205041].
}

\lref\VafaWitten{
  C.~Vafa and E.~Witten,
  Nucl.\ Phys.\ B {\bf 431}, 3 (1994)
  [arXiv:hep-th/9408074].
}

\noblackbox

\Title{\vbox{\baselineskip12pt\hbox{} }} {\vbox{\centerline{ Examples of
M5-Brane Elliptic Genera }}
}

\centerline{Davide Gaiotto and Xi Yin }
\smallskip
\centerline{Jefferson Physical Laboratory, Harvard University,
Cambridge, MA 02138} \vskip .6in \centerline{\bf Abstract} { We
determine the modified elliptic genus of an M5-brane wrapped on various one modulus
Calabi-Yau spaces,
using modular invariance together with some known Gopakumar-Vafa
invariants of small degrees. As a bonus, we find nontrivial
relations among Gopakumar-Vafa invariants of different degrees and
genera from modular invariance.
 } \vskip .3in

\Date{February 2007}

\listtoc\writetoc

\newsec{Introduction}

The modified elliptic genus of an M5-brane wrapped on a Calabi-Yau
space counts the degeneracies of D4-D2-D0 BPS black holes
\refs{\msw,\mms,\meg}. This partition function is modular invariant
and can be determined entirely by the knowledge of the degeneracies
of a finite number of states
\refs{\fareytail,\MinahanCT,\Estring,\deBoerVG,\KrausNB,\meg}. See
also \refs{\DabholkarDT,\denefwork,\adstop}. This was applied in
\meg\ to an M5-brane wrapped on
the hyperplane section of the quintic threefold. In this note, we
extend the result of \meg\ to some other Calabi-Yau spaces: the
sextic, octic, dectic in weighted projective spaces, as well as the
bicubic in ${\bf P}^5$.

The modified elliptic genus of an M5-brane wrapped on an ample divisor $P$ in Calabi-Yau space $X$ takes the form
\eqn\modell{ Z_{X,P}(\tau,\bar\tau,y^A) = \sum_{\delta \in \Lambda^*/\Lambda} Z_\delta(\tau) \Theta_{\Lambda+\delta}(\tau,
\bar\tau, y^A) } where $\Lambda\subset H^2(P,{\bf Z})$ is the image of
$$
\iota: H^2(X,{\bf Z})\hookrightarrow H^2(P,{\bf Z})
$$
$\Theta_{\Lambda+\delta}$ is the theta functions of the shifted
lattice $\Lambda+\delta$,\foot{There is an additional half integral
shift by $J/2$ due to a well known anomaly \refs{\moorecharges,
\freedwitten}.} \eqn\shith{ \Theta_{\Lambda+\delta}(\tau,\bar\tau,
y^A) = \sum_{\vec q\in \Lambda+\delta+{J\over 2}} (-)^{J\cdot
q}\exp\left[ -\pi i\tau {\vec q}^2 + \pi i(\tau-\bar\tau) {(J\cdot
q)^2\over J\cdot J}+2\pi iy\cdot q \right] } where $J$ is the
canonical class of $P$. $Z_\delta(\tau)$ are a set of holomorphic
modular vectors.\foot{When $P$ is not ample, there can be a
holomorphic anomaly in the $Z_\delta$'s \VafaWitten. This subtlety
does not appear in the examples we will be considering, and will be
ignored in this note. } $Z_{X,P}$ is expected to be a Jacobi form of
weight $(-{3\over 2},{1\over 2})$.

Our approach, as in \meg, is to determine $Z_{X,P}$ from the polar terms in the $q$-expansion of
$Z_\delta$'s. The latter involves the degeneracy of BPS D4-D2-D0 bound states with small charges, and can be determined
from geometric reasoning. In \meg, the geometric counting was ``naive" in that the authors did not take into account singularities
in the classical moduli space of the D4-D2-D0 bound states, which need to be resolved. There one needed to invoke
arguments based on the holographic dual of the M5-brane $(0,4)$ CFT to get the precise counting. In this note we
proposed a more refined counting solely based on the geometry, and we will see that it gives precisely the correct countings
that are consistent with the constraints imposed by modular invariance.

We will count D4-D2-D0 bound states with small charges by quantizing their classical moduli space.
The classical supersymmetric configuration of D4-D2-D0 system involves a hypersurface $P$ (hyperplane in our examples), with $U(1)$
fluxes represented by a type $(1,1)$ harmonic form $F$, together with $n$ point-like instantons (D0-branes). Up to the shift by $J/2$,
$F$ represents an integral class in $H^2(P,{\bf Z})$, and can be represented by an integral linear combination of holomorphic curves $C_i$
in $P$. It is most convenient to think of $C_i$'s as curves in $X$ that coincide with $P$. Note that two curves can be homologous in $X$ but
not homologous as classes in $P$. We will mostly think of the simple case when $C_i$'s are rigid curves.
In general they are counted by Gopakumar-Vafa invariants \refs{\gvi,\gv,\kkv}.\foot{When the curves have moduli, it is a priori not
obvious that Gopakumar-Vafa invariants are relevant for our counting. However, holography suggests that this should be the case \meg.
}
 One can then think of (a component of)
the classical moduli space as the space ${\cal M}$ of hyperplane $P$ that passes through a set of given curves $C_i$ as well as $n$ points in $X$.

With $C_i$'s rigid and fixed, ${\cal M}$ is essentially a projective space fibered over the space of $n$ points in $X$. The index that counts BPS states is given by the Euler characteristic (in a suitable sense) of ${\cal M}$, which is easy to evaluate on the smooth components of ${\cal M}$.
The naive description of ${\cal M}$ based on classical geometry has a lot of singularities. For example, wherever some of $n$ points coincide with one another, or coinciding with one of the curves $C_i$, the dimension of the fiber projective space jumps
and ${\cal M}$ is singular. Physically, such singularities can often be resolved by the nonabelian degrees of
freedom of the D-branes. We expect ${\cal M}$ to be fibered over a resolved space of $n$ points in $X$ (possibly the Hilbert scheme).

For example, when we have two points $p_1, p_2$ colliding in $X$, it is straightforward to resolve the moduli space, replacing the locus where $p_1,p_2$ coincide by the space of directions along which $p_2$ can approach $p_1$, namely a ${\bf P}^2$.
Similarly, if a point $p$ collides with a curve $C$, we will replace the locus in the moduli space where $p$ lies on $C$ by the space of possible directions $p$ can approach $C$ (a ${\bf P}^1$ worth of them) fibered over $C$. When three points collide, we can replace the locus where the three points coincide by the space of planes spanned by three points infinitesimally close to one another (a ${\bf P}^2$ worth of them).
The resolution of the moduli space is not so straightforward when more than three points collide. It can presumably be understood in terms of the
nonabelian dynamics of the D-branes. Fortunately we will not need them in the examples considered in this note.

In the following section we will compute the modified elliptic genus
for an M5-brane wrapped on the hyperplane section in the quintic in
${\bf P}^4$, sextic in ${\bf WP}_{2,1,1,1,1}$, octic in ${\bf
WP}_{4,1,1,1,1}$, dectic in ${\bf WP}_{5,2,1,1,1}$, and the bicubic
in ${\bf P}^5$. We will make use of the Gromov-Witten and
Gopakumar-Vafa invariants computed in \refs{\KlemmTX,\kkv} (more
complete results can be found in \crazy). The details of the modular
vectors involved are described in the appendix.

\newsec{The M5-brane elliptic genus on a number of Calabi-Yau spaces}

\subsec{The quintic in ${\bf P}^4$, revisited}

In this subsection we recall the result of \meg, but will recount
the degeneracies of BPS states of small charges from a refined
geometric picture. The modified elliptic genus takes the form
\eqn\quintcc{ Z_{X_5}(\tau,\bar\tau, y) = \sum_{i=0}^4 Z_i(\tau)
\Theta_i^{(5)}(\bar\tau, y) } where \eqn\tehtaf{
\Theta^{(m)}_k(\tau, y) \equiv \sum_{n\in {\bf Z}+\half+{k\over m}}
(-)^{mn} q^{{m\over 2}n^2} e^{2\pi i y mn} } and the $Z_i$'s are
given by \eqn\exactmod{
\eqalign{&Z_0(q)=q^{-{55\over24}}(5-800q+58500q^2
+5817125q^3+75474060100q^4+28096675153255q^5+\cdots) \cr &Z_1(q) =
Z_4(q)= q^{-{83\over 120}}(8625
-1138500q+3777474000q^2+3102750380125q^3+\cdots) \cr &Z_2(q)=Z_3(q)=
q^{{13\over 120}}
(-1218500+441969250q+953712511250q^2+217571250023750q^3+\cdots) }}
As a nontrivial check, the number of D4 bound to 2 D0-branes can be
counted by considering a hyperplane that passes through two points,
say $p_1,p_2$. When $p_1$ and $p_2$ are distinct, there is a ${\bf
P}^2$ worth of hyperplanes that pass through both points. When
$p_1,p_2$ collide, we need to resolve the moduli space and take into
account the directions $p_2$ can approache $p_1$. This amounts to
replace the locus in the moduli space where the two points collide
by a ${\bf P}^2$. In this case, we shall require not only $p_1=p_2$
lie in the hyperplane, but the vector determined by the direction
along which $p_2$ approaches $p_1$ lie in the hyperplane as well.
This again determines a ${\bf P}^2$ worth of hyperplanes. The
counting is
$$
(-200)\cdot(-201)/2\cdot \chi({\bf P}^2) +(-200)\cdot \chi({\bf
P}^2)\cdot \chi({\bf P}^2) = 58500,
$$
which indeed agrees with the corresponding coefficient in $Z_0$, predicted by modular invariance.

Next let us consider a D4 with one unit of flux, and bound to one
extra D0. This is counted by a hyperplane that passes through a
degree 1 rational curve $C_1$, as well as an extra point $p$. When
$p$ does not lie on $C_1$, $p$ and $C_1$ determine a ${\bf P}^1$
worth of hyperplanes. When $p$ collides with $C_1$, the moduli space
is resolved so that it contains the space of directions along which
$p$ can approach $C_1$ at any given point, which is another ${\bf
P}^1$. So that counting is
$$
2875\cdot (-200-\chi(C_1))\cdot \chi({\bf P}^1) +2875\cdot
\chi(C_1)\cdot\chi({\bf P}^1)\cdot\chi({\bf P}^1)=-1138500,
$$
precisely agreeing with the corresponding coefficient in $Z_1$.

Now consider a D4 with two units of fluxes and D0-brane charge one
more than the minimal value (the second coefficient in the
$q$-expansion of $Z_2$). The counting receives three contributions:
a hyperplane that passes through a degree 2 rational curve $C_2$ and
a point $p$, with the flux being $F=C_2$; a hyperplane that passes
through two distinct degree 1 rational curves $C_1$ and $C_1'$, with
the flux being $F=C_1+C_1'$; or a hyperplane that passes through a
degree 3 rational curve $C_3$, with the flux being $F=J-C_3$. In the
first case, we again need to resolve the locus of the moduli space
where $p$ collides with $C_2$, as before. There is also an extra
minus sign one needs to take into account as in \meg.\foot{We do not
know how to understand this directly from quantizing the classical
moduli space. This is not a contradiction since disconnected branches of the
moduli space can contribute with different signs. This sign was determined in \meg\ 
from the fermion number of the
wrapped M2-brane in the holographic dual.} Using the well known
Gromov-Witten invariants of degree 1,2,3, the counting is
$$
(-609250)\cdot \left[ -200-\chi(C_2)+ \chi(C_2)\cdot\chi({\bf
P}^1)\right]+ {2875\choose 2}+317206375=441969250,
$$
which again agrees with the prediction from modular invariance.

A more difficult example is a D4 bound to 3 D0's. There are
essentially two kinds of contributions: a hyperplane that passes through two
different degree 1 rational curves $C_1$ and $C_1'$, with the flux
being $F=C_1-C_1'$; or a hyperplane that passes through three points
$p_1,p_2,p_3$. The contribution from the first case is
straightforward: $C_1$ and $C_1'$ complete determines a hyperplane.
The second case is more subtle due to the different configurations
of the three points. Naively, there are five different situations
one must consider:

\noindent (a) $p_1,p_2,p_3$ are distinct and are not aligned in the
ambient ${\bf P}^4$. The three points determine a ${\bf P}^1$ worth
of hyperplanes.

\noindent (b) $p_1,p_2,p_3$ are distinct and lie on a line $L$ in
${\bf P}^4$ (which intersects the quintic at five points). $L$
determines a ${\bf P}^2$ (as opposed to a ${\bf P}^1$) worth of
hyperplanes.

\noindent (c) $p_1=p_2\not=p_3$. When resolving the moduli space by
taking into account of the direction $\overline{p_1p_2}$, $p_3$ does
not lie on the line determined by $\overline{p_1p_2}$ in the ambient
${\bf P}^4$.

\noindent (d) $p_1=p_2\not=p_3$. $p_3$ is one of the remaining 3
intersections of the line determined by $\overline{p_1p_2}$ with the
quintic in the ${\bf P}^4$.

\noindent (e) $p_1=p_2=p_3$. Resolving the moduli space replaces the
point by a ${\bf P}^2$ worth of planes spanned by three close by
points. Each such plane determines a ${\bf P}^1$ worth of
hyperplanes.

Putting these together, we get the counting
$$
\eqalign{& 2875\cdot 2874 + {(-200)\cdot (-201)\cdot (-200-5)\over
6}\chi({\bf P}^1)+{(-200)\cdot (-201)\cdot 3\over 6}\chi({\bf P}^2)
\cr & + (-200)\cdot (-200-4)\cdot\chi({\bf P}^2)\cdot \chi({\bf
P}^1) + (-200)\cdot 3\cdot \chi({\bf P}^2) \cdot\chi({\bf P}^2) +
(-200)\cdot \chi({\bf P}^2)\cdot \chi({\bf P}^1) \cr &= 5814250 =
5817125-2875. }
$$
It is striking yet puzzling that the result differs from the
prediction from modular invariance, 5817125, by $-2875$ (recall that 2875 is the number
of degree 1 rational curves in the quintic). In the
above counting we have ignored the more complicated situation where
the points $p_1,p_2,p_3$ lie on a degree 1 curve $C_1$ (as opposed
to a generic line $L$). The corrections one obtain by taking into
account such configurations will presumably be a multiple of 2875.
We do not understand why the multiplicity is ``1", which we will
leave to future investigation.

In summary, we found remarkable agreement of the modified elliptic
genus with the proposed geometric counting by resolving the
singularities of the moduli space.

\subsec{Degree 6 hypersurface in ${\bf WP}_{(2,1,1,1,1)}$}

The Calabi-Yau 3-fold $X_6$ is defined as the hypersurface \eqn\aa{
x_1^3+x_1 f_4(x_2,x_3,x_4,x_5)+f_6(x_2,x_3,x_4,x_5)=0 } in the
weighted projective space ${\bf WP}_{(2,1,1,1,1)}$, where $f_4$ and
$f_6$ are polynomials of homogeneous degree 4 and 6 in
$x_2,\cdots,x_5$. We will assume that $f_4,f_6$ are generic and
$X_6$ is smooth. The choice of complex structure is not essential
for our purpose. $X_6$ has $h^{1,1}=1$, $h^{2,1}=103$, $\chi=-204$,
$c_2=14h$, $h$ being generator of $H^4(X_6,{\bf Z})$. The hyperplane
section $P$ has $6D=P\cdot P\cdot P=3$, $c_2\cdot P=42$. The
M5-brane $(0,4)$ CFT has left and right central charges
$$
c_L=6D+c_2\cdot P=45, ~~~ c_R = 6D+\half c_2\cdot P=24.
$$
The modified elliptic genus takes the form \eqn\sexticsg{
Z_{X_6}(\tau,\bar\tau, y) = \sum_{i=0}^2 Z_i(\tau)
\Theta_i^{(3)}(\bar\tau, y) } where the $\Theta_i^{(3)}$'s are
defined as in \tehtaf, and $Z_1=Z_2$. A direct counting from
geometry gives the polar terms \eqn\oapsf{ \eqalign{ &Z_0(\tau) =
q^{-{45\over 24}} (4+3\cdot (-204)q+\cdots) \cr & Z_1(\tau) =
q^{-{45\over 24}-{1\over 3}+2}(2\cdot 7884+\cdots) } } where
$\cdots$ are non-polar terms, of higher orders in $q$. This
determines the modified elliptic genus by modular invariance. We
will leave the details of the modular forms to the appendix, and
write the first few terms in the $q$-expansion here \eqn\asfqqq{
\eqalign{ &Z_0(\tau) = q^{-{45\over 24}} (4 - 612 q + 40392 q^2 -
    146464860 q^3 - 66864926808 q^4 -
    8105177463840 q^5 \cr &~~~- 503852503057596 q^6 -
    20190917119833144 q^7 - 587565090039987648 q^8+\cdots),  \cr
&Z_1(\tau)=Z_2(\tau)=q^{-{5\over 24}} (15768 - 7621020 q -
10739279916 q^2 -
    1794352963536 q^3 \cr &~~~- 134622976939812 q^4 -
    6141990299963544 q^5 - 196926747589177416 q^6 +\cdots). } }
Let us make a few checks. The number of D4 bound to 2 D0's can be
counted directly from the geometry. Naively, by resolving the moduli space
of a hyperplane passing through two points $p_1,p_2$ as before we get
$$
(-204)(-205)/2 \cdot\chi({\bf P}^1)+ (-204)\cdot\chi({\bf P}^2)\cdot \chi({\bf P}^1)=40596
$$
which differs from the expected answer 40392 by 204.
The reason for this discrepancy is a simple geometric fact: hyperplane
sections of the sextic are defined by linear equations in the four degree 1 variables $x_2, \cdots, x_5$ only.
Given a point $p_1$, all hyperplanes through $p_2$ will also pass through two other points in the
sextic with the same $x_2, \cdots, x_5$ coordinates but with different $x_1$ coordinates.
If $p_2$ is one of these two points, it will not constrain the hyperplane any
further, and hence there is a ${\bf P}^2$, instead of a ${\bf P}^1$, worth of hyperplanes through $p_1,p_2$.
This gives a correction $(-204)\cdot2/2$ to the degeneracy. In the end we get $40596-204=40392$
which precisely agrees with \asfqqq.

The number of D4 bound to 1 D2 and 1 D0 can be counted directly:
$$
-6028452+(-204-2)\cdot 7884 + 2\cdot 2\cdot 7884=-7621020
$$
which again exactly matches the predicted answer in $Z_1(\tau)$.

\subsec{Degree 8 hypersurface in ${\bf WP}_{(4,1,1,1,1)}$}

The Calabi-Yau 3-fold $X_8$ is defined as the hypersurface \eqn\aa{
x_1^2+f_8(x_2,x_3,x_4,x_5)=0 } in the weighted projective space
${\bf WP}_{(4,1,1,1,1)}$, with $f_8$ a polynomial of homogeneous
degree 8. $X_8$ has $h^{1,1}=1$, $h^{2,1}=149$, $\chi=-296$,
$c_2=22h$. The hyperplane section $P$ has $6D=P\cdot P\cdot P=2$,
$c_2\cdot P=44$. The M5-brane $(0,4)$ CFT has central charges
$$
c_L=6D+c_2\cdot P=46, ~~~ c_R = 6D+\half c_2\cdot P=24.
$$
The modified elliptic genus takes the form \eqn\assb{
Z_{X_8}(\tau,\bar\tau,y) = Z_0(\tau) \theta_2(2\bar\tau,2y) -
Z_1(\tau) \theta_3(2\bar\tau,2y) } Direct counting from geometry
gives the polar terms of $Z_{0,1}(\tau)$, \eqn\fiss{ \eqalign{&
Z_0(\tau) = q^{-{46\over 24}}(4+3\cdot(-296) q+\cdots)  \cr &
Z_1(\tau) = q^{-{46\over 2}-{1\over 4}+2} (2\cdot 29504+\cdots) } }
These determine the modified elliptic genus completely. We will
leave the details of the modular forms to the appendix, and write
the first few terms in the $q$-expansion here \eqn\modfas{ \eqalign{
&Z_0(\tau) = q^{-{46\over 12}} (4 - 888 q + 86140 q^2 -
    132940136 q^3 - 86849300500 q^4 \cr &~~~-
    11756367847000 q^5 - 787670811260144 q^6 -
    33531427162546608 q^7 +\cdots) \cr & Z_1(\tau) = q^{-{1\over 6}}
(59008 - 8615168 q- 21430302976 q^2 -
    3736977423872 q^3 \cr &~~~- 289181439668352 q^4 -
    13588569634434304 q^5 - 448400041603851008 q^6 +\cdots) } }
Let us make a few checks. The direct counting of D4 bound to D2 and
a D0 gives the $q^{5\over6}$ coefficient of $Z_1(q)$
$$
29504\cdot (-296-2) + 2\cdot \chi({\bf P}^2) 29504 = -8615168
$$
which exactly matches the prediction from modularity. A direct
counting of D4 bound to 2 D0 gives
$$
2 (-296)\cdot (-297)/2+2 \chi({\bf P}^2) (-296)=86136
$$
which differs from the expected answer 86140 in $Z_0(q)$ by
4. Similar to the case of the sextic, we expect a correction due to the fact that
a hyperplane though $p_1$ also necessarily passes through one other point $p_2$ with
the same $x_2, \cdots, x_5$ coordinates as $p_1$.
This would give a correction $(-296)\cdot 1/2=-148$ to the degeneracy. This brings the discrepancy
with the expected answer from \modfas\ to
$152$. This does not necessarily imply a failure of the geometric counting, since there are potentially
holomorphic curves that can contribute to the number of BPS states with the same charges.
A degree $d=2m$ genus $g$ curve $C$ in $P$ has self-intersection $C \cdot C = 2g-2-2m$ and
turning on the flux $F=C-mJ$ would
induce D0 charge $-{(C-mJ)^2 \over 2} =m^2+ m+1-g$. For example, any $d=2,g=1$ or $d=4,g=5$ curve that
lies on a hyperplane in $X_8$ could contribute to the degeneracy and they might account for the above
 discrepancy. We will leave this point to future investigation.

\subsec{Degree 10 hypersurface in ${\bf WP}_{(5,2,1,1,1)}$}

$X_{10}$ is the hypersurface defined by a polynomial of homogeneous
degree 10 in the weighted projective space ${\bf WP}_{(5,2,1,1,1)}$.
It has $h^{1,1}=1$, $h^{2,1}=145$, $\chi=-288$, $c_2=34h$. The
hyperplane section $P$ has $6D=P\cdot P\cdot P=1$, $c_2\cdot P=34$.
Note that $P$ is defined by a linear equation in $x_3,x_4,x_5$ only.
The $(0,4)$ CFT has central charges
$$
c_L=6D+c_2\cdot P=35, ~~~ c_R = 6D+\half c_2\cdot P=18.
$$
A straightforward counting of D4-D0 bound state with D0 charge $0,1$
determines the first two terms in the modified elliptic genus
$$
Z_{X_{10}}(\tau,\bar\tau,y) = q^{-{35\over 24}}(3+2\cdot (-288)
q+\cdots) \theta_1(\bar\tau,y)
$$
Requiring that $Z_{X_{10}}$ is a Jacobi form of weight $(-{3\over
2},{1\over 2})$ then determines it to be \eqn\resxt{\eqalign{
&Z_{X_{10}}(\tau,\bar\tau,y) = \eta(\tau)^{-35} {541 E_4(\tau)^4 +
1187 E_4(\tau) E_6(\tau)^2\over 576}\theta_1(\bar\tau,y) \cr &=
q^{-{35\over 24}}(3 - 576 q + 271704 q^2 +
    206401533 q^3 + 21593767647 q^4 +
    1054723755951 q^5+\cdots) \theta_1(\bar\tau,y) }}
A naive direct counting of D4 bound to 2 D0's give
$$
(-288)\cdot (-289)/2+(-288) \chi({\bf P}^2) +231200 = 271952
$$
which is 248 more than the value 271704 predicted by modular
invariance. Now a hyperplane through one point $p_1$ will also contain a whole curve
with the same $x_3,x_4,x_5$ coordinates. There is again a correction to the degeneracy when $p_2$ lies
on this curve, which is more subtle since the curve may degenerate depending on the $x_3,x_4,x_5$ coordinates.
And furthermore, a genus $g$ degree 1 curve $C_g$ would have self-intersection $2g-3$ and the flux $C_g-J$ would carry
D0 charge $2-g$. Such curves may contribute if they lie in a hyperplane section. A careful analysis of these contributions
is beyond this note.

\subsec{Bicubic in ${\bf P}^5$}

$X_{3,3}$ is defined by \eqn\aa{ P_3(X)=Q_3(X)=0 } in ${\bf P}^5$,
where $P$ and $Q$ are generic cubic polynomials. $X_{3,3}$ has
$h^{1,1}=1$, $h^{2,1}=73$, $\chi=-144$, $c_2=6h$. The hyperplane
section $P$ has $6D=P\cdot P\cdot P=9$, $c_2\cdot P=54$. The $(0,4)$
CFT has
$$
c_L=6D+c_2\cdot P=63, ~~~ c_R = 6D+\half c_2\cdot P=36.
$$
The modified elliptic genus has the form \eqn\asfbic{
Z_{X_{3,3}}(\tau,\bar\tau,y) = \sum_{i=0}^8 Z_i(\tau)
\Theta_i^{(9)}(\bar\tau, y) } where $\Theta_i^{(m)}$ are defined as
before. There is also the relation $Z_i=Z_{9-i}$. By direct counting
from geometry we can determine the first few terms in the
$q$-expansion of the $Z_i$'s \eqn\poscic{ \eqalign{ &Z_0(\tau) =
q^{-{63\over 24}}(6+5\cdot (-144)q+(?)q^2+\cdots) \cr &Z_1(\tau) =
q^{-{77\over 72}}(4\cdot 1053+ 3\cdot 1053\cdot (-144+2)q+\cdots)
\cr &Z_2(\tau) = q^{-{29\over 72}}(-3\cdot 52812+\cdots) \cr &
Z_3(\tau)=q^{-{5\over 8}}(3\cdot(-3402)+\cdots) \cr &Z_4(\tau) =
q^{19\over 72}(2\cdot 5520393+\cdots)} } We did not try to determine
the (?) coefficient in $Z_0(\tau)$ because of the potential
ambiguity in the counting from geometry. However we can count the
first (non-polar) coefficient in $Z_4(\tau)$ from degree 4 genus 1
curves in $P$, and together with the other polar coefficients they
determine the modified elliptic genus completely.

The details of determining the modular forms are left to the
appendix.\foot{As explained in the end of appendix A.2, this modular form has an unexpected
feature, suggesting a yet uncovered mysterious relation among the polar coefficients.} The first few terms in the $q$-expansion of the answer are
given by \eqn\poscicans{ \eqalign{ &Z_0(\tau) = q^{-{63\over
24}}(6-720 q+40032 q^2 + 678474 q^3 -
    30885198768 q^4 - 35708825468142 q^5\cr &~~~ -
    9448626104689554 q^6 - 1170512868283650738 q^7 -
    88016808046791466314 q^8+\cdots) \cr &Z_1(\tau) = q^{-{77\over 72}}(4212
 - 448578 q - 374980104 q^2 -
    2020724648442 q^3 - 890559631782378 q^4 \cr &~~~-
    147810582092632410 q^5 - 13583665805416442478 q^6 -
    823655461162634305794 q^7 +\cdots) \cr &Z_2(\tau) = q^{-{29\over
72}}(-158436 + 12471246 q -
    174600085086 q^2 - 134299669045176 q^3 \cr &~~~-
    29070064587874050 q^4 - 3172859337263652090 q^5 -
    218000892267121506858 q^6+\cdots) \cr & Z_3(\tau)=q^{-{5\over
8}}(-10206 + 13828428 q -
    24425287884 q^2 - 35338801262184 q^3\cr &~~~ -
    9438086780879238 q^4 - 1170314443959539166 q^5 -
    88014001223404540188 q^6 +\cdots) \cr &Z_4(\tau) = q^{19\over 72}
(11040786 - 6769752552 q -
    17629606262268 q^2 - 5304774206609694 q^3 \cr &~~~-
    704390403350490336 q^4 - 55554435778447164564 q^5 +\cdots)} }
We can make one highly nontrivial check: the second coefficient in
$Z_3(\tau)$, 13828428, is the number of D4 bound to 3 D2's with an
extra D0-brane charge. From the geometric picture this comes from a
degree 3 genus 0 curve $C_{3,0}$ lying in $P$, as well as a degree 3
genus 1 curve $C_{3,1}$ together with a pointlike instanton in $P$.
The counting (using known results of Gopakumar-Vafa invariants that
count $C_{3,0}$ and $C_{3,1}$) is
$$
2\cdot 6424326+2\cdot (-3402)\cdot (-144) = 13828428
$$
which precisely agrees with the prediction from modular invariance.

\subsec{Quadriconic in ${\bf P}^7$}

$X_{2,2,2,2}$ is defined by \eqn\aa{ P_2(X)=Q_2(X)=R_2(X)=S_2(X)=0 }
in ${\bf P}^7$, where $P,Q,R,S$ are generic quadratic polynomials.
$X_{2,2,2,2}$ has $h^{1,1}=1$, $h^{2,1}=65$, $\chi=-128$, $c_2=4h$.
The hyperplane section $P$ has $6D=P\cdot P\cdot P=16$, $c_2\cdot
P=64$. The $(0,4)$ CFT has
$$
c_L=6D+c_2\cdot P=80, ~~~ c_R = 6D+\half c_2\cdot P=48.
$$
The modular forms involved are more complicated and the determination of the
D4-brane partition function in this case is left as a fun exercise for the reader.

\bigskip

\centerline{\bf Acknowledgement} We are grateful to Vicent Bouchard, Daniel Jafferis, Greg Moore and
Andy Strominger for useful discussions. DG is
supported in part by DOE grant DE-FG02-91ER40654. XY is supported by a Junior Fellowship
from the Harvard Society of Fellows.

\appendix{A}{The details of modular forms}

\subsec{The method of generating modular representations}

In this section we describe an algorithm to find many independent
modular vectors to construct a basis of the relevant modular
representation, the number of elements in the basis being the number
of allowed polar terms in the $q$-expansion of the modular vector.

We can start with a vector $(\chi_i^w(\tau))_{i=0,\cdots,m-1}$ transforming in a particular
$m$-dimensional modular representation with weight $w$ (half integer in general),
and obtain a weight $w+2$ vector in the same representation by
\eqn\denss{ {\cal D}_2(\chi_i^{w})(\tau) := {1\over 2\pi
i}\eta(\tau)^{2w}\partial_\tau (\eta(\tau)^{-2w}\chi_i^w(\tau)) }
One can repeat this procedure and get modular vector of weight
$w+2n$. The modular vector obtained this way (for $n>1$) are not
necessarily the same as $\chi^w_i(\tau)$ multiplied by entire
holomorphic modular forms (polynomials in $E_4, E_6$).

The first step is to find ``seeding" modular forms
$\chi_i^w(\tau)$ that transform in the same representation as
$\Theta_{1,i}^{m}(\tau,y)$. Here the theta functions are defined
as \eqn\tehtaaa{ \eqalign{ &\Theta_{1,k}^{m}(\tau,y) = \sum_{n\in
{\bf Z}+\half+{k\over m}} (-)^{mn}q^{{m\over 2}n^2}z^{mn} \cr
&\Theta_{2,k}^{m}(\tau,y) = \sum_{n\in {\bf Z}+\half+{k\over m}}
q^{{m\over 2}n^2}z^{mn} \cr &\Theta_{3,k}^{m}(\tau,y) = \sum_{n\in
{\bf Z}+{k\over m}} q^{{m\over 2}n^2}z^{mn} \cr
&\Theta_{4,k}^{m}(\tau,y) = \sum_{n\in {\bf Z}+{k\over m}}
(-)^{mn}q^{{m\over 2}n^2}z^{mn} } } where $z=e^{2\pi iy}$,
$k=0,\cdots,m-1$. These are the usual Jacobi theta functions for
$m=1$, $k=0$. Only the $\Theta_{1,k}^m(\tau,y)$'s form an
$m$-dimensional modular representation by themselves. However they
vanish at $y=0$, and we need to come up with the $\chi_i^w(\tau)$'s that
transform in the same way with some weight $w$.

A good set of seeding modular forms is \eqn\seedinga{ \eqalign{ &
\chi_i^{m,4l-{m-1\over 2}}(\tau) = \theta_3(\tau)^{8l-m}
\Theta_{3,i}^m(\tau)+\theta_4(\tau)^{8l-m}
\Theta_{4,i}^m(\tau)+\theta_2(\tau)^{8l-m}
\Theta_{2,i}^m(\tau),~~~~~~~~~m~{\rm odd}; \cr
&\chi_i^{m,4l-{m-1\over 2}}(\tau) = \theta_3(\tau)^{8l-m}
\Theta_{3,i}^m(\tau)+(-)^k\theta_4(\tau)^{8l-m}
\Theta_{3,i}^m(\tau)+\theta_2(\tau)^{8l-m}
\Theta_{2,i}^m(\tau),~~~m~{\rm even}.} } Here the first superscript
of $\chi$ indicates its modular representation, i.e. that of
$\Theta_1^m$; the second superscript indicates its modular weight,
and the subscript is the index for the modular vector. The choice of
$\chi$ is motivated by the $S$ and $T$ transformation of the
$\Theta_{i,k}^m$ of the form
$$
\eqalign{ & \Theta_2^m\longleftrightarrow^{\!\!\!\!\!\!\!\! S}~\,
\Theta_3^m\longleftrightarrow^{\!\!\!\!\!\!\!\!
T}~\,\Theta_4^m,~~~~m~{\rm odd} \cr
&\Theta_2^m\longleftrightarrow^{\!\!\!\!\!\!\!\! S}~\,
\Theta_3^m\longleftrightarrow^{\!\!\!\!\!\!\!\!
T}~\,(-)^k\Theta_3^m,~~~~m~{\rm even}}
$$
relative to the modular transform of $\Theta_1^m$. There are in
general more possible seeding modular forms, but \seedinga\ appears
to suffice for our purpose.

\subsec{The results}

The modified elliptic genus of an M5-brane wrapped on the hyperplane
section in the octic in ${\bf WP}_{4,1,1,1,1}$ is \eqn\osctc{
\eqalign{ Z_{X_{8}}(\tau,\bar\tau,y) &= {1\over
63}\eta^{-46}\sum_{k=0,1}(77 E_4^3 E_6 \chi_k^{2,{7\over
2}}-19278E_6\Delta \chi_k^{2,{7\over 2}}-168 E_4^4 {\cal
D}_2\chi_k^{2,{7\over 2}}\cr &+245808E_4\Delta {\cal
D}_2\chi_k^{2,{7\over 2}})\Theta_{1,k}^2(\bar\tau,y) }} where $E_4,
E_6, \Delta\equiv \eta^{24}$ and $\chi$ are understood to be
functions of $\tau$.

The modified elliptic genus of an M5-brane wrapped on the hyperplane
section in the sextic in ${\bf WP}_{2,1,1,1,1}$ is \eqn\sexctc{
\eqalign{ Z_{X_{6}}(\tau,\bar\tau,y) &= {1\over
4}\eta^{-45}\sum_{k=0}^2(5E_4^3 E_6 \chi_k^{3,3}-1344 E_6\Delta
\chi_k^{3,3}-12 E_4^4 {\cal D}_2\chi_k^{3,3} \cr & + 15360 E_4\Delta
{\cal D}_2\chi_k^{3,3} )\Theta_{1,k}^3(\bar\tau,y) }}

The modified elliptic genus of an M5-brane wrapped on the hyperplane
section in the bicubic in ${\bf P}^5$ is \eqn\biccans{ \eqalign{
&Z_{X_{3,3}}(\tau,\bar\tau,y)  = {1\over 698880}\eta^{-63}
\sum_{k=0}^8 \left[ (-174720 E_4^5 E_6 - 39370048 E_4^2 E_6 \Delta)
\chi^{9,4}_k \right. \cr &~~~+ (704340 E_4^6 - 1205445441 E_4^3
\Delta + 143587676160 \Delta^2) {\cal D}_2\chi^{9,4}_k\cr
&~~~+(176904 E_4^4 E_6 - 952935930 E_4 E_6 \Delta){\cal
D}_2^2\chi_k^{9,4} +(-6368544 E_4^5 + 2752749684 E_4^2\Delta){\cal
D}_2^3\chi_k^{9,4} \cr & ~~~\left. + (19105632 E_4^3 E_6 -
3794532480 E_6 \Delta) {\cal D}_2^4 \chi_k^{9,4} + 7233791184 E_4
\Delta {\cal D}_2^5 \chi_k^{9,4} \right] \Theta_k^{9}(\bar\tau,y) }
} Here we constructed in fact one fewer basis modular vectors than
all possible polar terms, nevertheless we seem to be lucky enough to
match all the polar terms obtained from geometric counting. This
suggests that there might be a hidden relation among the polar terms
that is not determined by modular invariance.

\listrefs

\end